\documentstyle[ltwol,psfig]{article}

\def\Journal#1#2#3#4{{#1} {\bf #2}, #3 (#4)}

\def\NPB{{\em Nucl. Phys.} B}

\def\PRL{\em Phys. Rev. Lett.}
\def\PRD{{\em Phys. Rev.} D}

\def\ra{\rightarrow}

\def\be{\begin{equation}}
\def\ee{\end{equation}}
\def\bea{\begin{eqnarray}}
\def\eea{\end{eqnarray}}
\def\CP{{\em CP\/}}
\begin{document}

\title{{\it CP}\/ Violation Search in $\tau\rightarrow \pi\pi^0\nu_{\tau}$ Decays}

\author{Thomas Coan}

\address{Physics Department, Southern Methodist University, 
Dallas, TX 75275, USA\\E-mail: coan@mail.physics.smu.edu}


\twocolumn[\maketitle\abstracts{ Using a sample of 12.2 million
$\tau$-lepton pairs produced by $e^+e^-$ annihilation at
$\sqrt{s}\sim 10.6\,$GeV and collected by the CLEO detector, we search
for and set limits on $CP$ violation in $\tau$-lepton decays. For each
event, we require that both $\tau$-leptons decay via the mode
$\tau\ra \pi\pi^0\nu$.  The search is performed within the
context of a multi-Higgs Doublet Model and the imaginary part of the
coupling constant parameterizing the non-Standard Model diagram
leading to $CP$ violation is constrained to be $-0.046 <
\Im{(\Lambda}) < 0.022$ at 90\% CL.  The novel search technique is of
general utility.}]

\section{Introduction}

{\em CP} violation has long been observed in the kaon system and is
expected soon to be seen unambiguously in the $B$-meson system.
Searches for {\em CP\/} violation in leptonic decays have been less
extensive. Although {\em CP\/} non-conservation in such decays is
forbidden in the Standard Model (SM), many extensions~\cite{ein} to
the SM incorporate it. Multi-Higgs Doublet models
(MHDM)~\cite{gross1}${}^{,\,}$\cite{albright} are perhaps the best
known.  This report describes a novel search technique for $CP$
violation performed with tau decays and for a specific MHDM. The
technique is general and no feature of it restricts it to our
particular choice of MHDM or final state.

With the CLEO detector, we study $e^+e^- \rightarrow \tau^+\tau^-$
events produced at CESR at a center-of-mass energy in the vicinity of
the $\Upsilon(4S)$ resonance. We select only those events where each
$\tau$ lepton decays in the mode $\tau\rightarrow \pi\pi^0\nu_{\tau}$.
Interference between the SM process mediated by $W$-boson exchange and
a non-SM process mediated by scalar boson exchange could lead to $CP$
violating effects.  For each of our events, we construct a $CP$
sensitive variable for which a non-zero value is evidence of $CP$
violation.

\begin{figure}[htbp]
\center
\psfig{figure=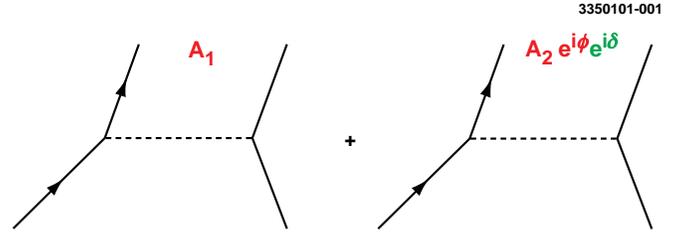,width=3.5in}
\caption{Generic Feynman diagrams for two transition amplitudes
that contain CP-even and CP-odd relative phases.}
\label{fig:interfere}
\end{figure}

\CP\/ violation is essentially an interference effect between two (or
more) transition amplitudes that have a relative phase difference for
both the strong and weak phases. The relative strong phase remains
invariant under {\em CP\/} conjugation while the relative weak phase
changes sign.  Figure~\ref{fig:interfere} shows the basic idea, where
$\delta$ and $\phi$ are the relative strong and weak phases,
respectively, and the overall transition amplitude $A=A_1 + A_2{\rm
e}^{i\phi}{\rm e}^{i\delta}$ for a process leads to its corresponding
probability density $P$:

\be
\begin{array} {rcl}

 P\propto |A^2| & = & (A_1 + A_2{\rm e}^{i\phi}{\rm e}^{i\delta})
(A_1 + A_2{\rm e}^{-i\phi}{\rm e}^{-i\delta})\\
& = & A_1^2 +A_2^2 + 2 A_1A_2\cos{\phi}\cos{\delta}\\
& &  - 2A_1A_2\sin{\delta}\sin{\phi}

\end{array}
\label{eq:a2}
\ee

\noindent Note that the last term contains the CP-odd
factor $\sin{\delta}\sin{\phi}$, which changes sign for
the {\em CP}-conjugate process, leading to difference
in probability densities between a particular transition
and its {\em CP\/} conjugate.

\section{Model Choice}

\begin{figure}[htbp]
\center
\psfig{figure=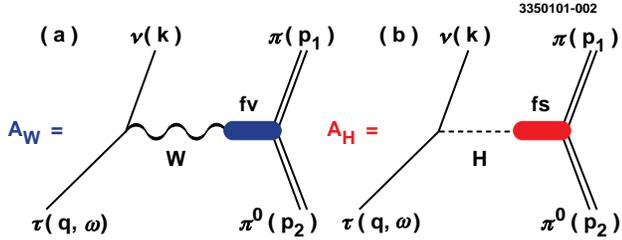,width=3.25in}
\caption{Feynman diagrams for a) the SM amplitude mediated by
$W$ exchange and b) the non-SM amplitude mediated by scalar exchange.}
\label{fig:3hdm}
\end{figure}

To make the foregoing generalities specific and to illustrate the
overall search technique, we select a specific MHDM model: a 3 Higgs
Doublet model~\cite{weinberg}. The relevant Feynman diagrams for this
model describing $\tau\rightarrow\pi\pi^0\nu_{\tau}$ are shown in
figure~\ref{fig:3hdm}. The SM diagram mediated by a $W$ boson is shown
on the left while the non-SM diagram mediated by a charged Higgs is
shown on the right. For the SM diagram, the vector form factor that
characterizes the final state strong interaction among quarks contains
a {\em CP}-even phase. The non-SM diagram contains a scalar form
factor $f_s$, also with a $CP$-even strong phase. The choice of $f_s$
is not unique so we consider three choices: $f_s=1$, ${f_s=\rm
BW[a_0(980)]}$, and ${f_s=\rm BW[a_0(1450)]}$. The term BW specifies a
Breit-Wigner shape and ${\rm a_0}$ indicates a particular intermediate
meson resonance.

In this model the second amplitude contains an overall complex
coupling $\Lambda$ between the charged Higgs H and the final state,
where $\Lambda$ is a function of the $H$ coupling to quarks and
leptons. Because $\Lambda$ is complex, it introduces a $CP$ odd phase.
Since we have two interfering diagrams with different relative strong
and weak phases, {\em CP\/} violation is possible.

\section{Choice of Optimal {\em CP}-sensitive Variable}

In general, we can write the $\tau$ production and decay probability
density $P$ as the sum of two terms, one invariant to CP conjugation
of the underlying Feynman amplitudes, $P_{even}$, and another that
changes under CP conjugation: $P= P_{even} + P_{odd}$. Both $P_{even}$
and $P_{odd}$ are functions of the form factors, the complex coupling
$\Lambda$ and kinematic quantities. To search for CP-violation we
select the CP-sensitive (i.e. CP odd) observable with the greatest
statistical significance and compute it for each event. Such a
variable~\cite{soni}${}^{,\,}$\cite{gunion} has been constructed with
different searches in mind by other workers:

\be
\xi = {{\rm P_{odd}\over {\rm P_{even}}}}
\ee

$\xi$ is a complicated function, and although it is maximally
sensitive to CP violation, has no immediate intuitive appeal. Averaged
over the data set, $\langle \xi\rangle \neq 0$ implies {\em CP\/}
violation. It can be shown~\cite{cleo} that

 \be
\begin{array}{rcl}
\langle \xi \rangle & = & Im(\Lambda)\int {{\rm P^2_{odd}}\over {\rm P_{even}}}\,dLips\\[2.mm]

& = & c_1\lambda + c_3\lambda^3 + c_5\lambda^5 + \ldots ,

\end{array}
\label{eq:lambda}
\ee

\noindent where $\lambda \equiv Im(\Lambda)$ and the $c_i$ are
constants.

\begin{figure}[htbp]
\center
\psfig{figure=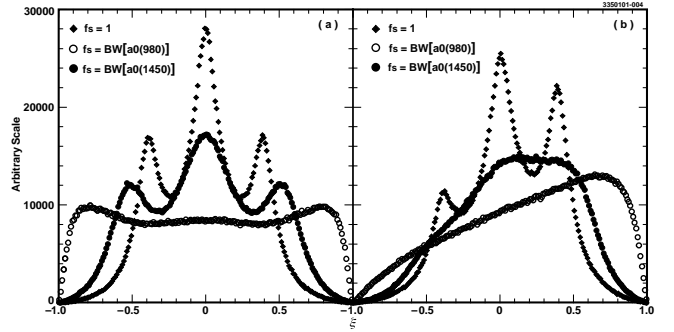,width=3.50in}
\caption{Monte Carlo distributions for the
optimal {\em CP}-sensitive $\xi$ for the case a) of no {\em CP\/} violation
and b) of `maximal' {\em CP\/} violation. See text for discussion.}
\label{fig:xi}
\end{figure}

The Monte Carlo $\xi$ distribution is shown in figure~\ref{fig:xi} for
the two limits of the model. The left-hand side is for no {\em CP\/}
violation and the right-hand side is for the `maximal' case where
$Im(\Lambda)$ has a value equal to the SM coupling. The three
distributions in each plot correspond to the three choices of the
scalar form factor and the structure in each distribution is due to
the resonant structure of the vector and scalar form factors. A non-zero
value of $\langle \xi\rangle$ for the {\em CP\/} violation case
is apparent.

\section{Experimental Details}

\subsection{Event Selection and Background Estimates}

To calculate a value of $\xi$ for each event it is necessary to know
the direction of the final state neutrinos to determine the direction
of the final state $\tau$-leptons. For those events where each $\tau$
decays semi-hadronically, energy and momentum conservation can be used
to determine each $\tau$ direction up to a two-fold
ambiguity~\cite{balest}.  Since we have no way of knowing the correct
solution for an event, we sum the $\xi$ distributions corresponding to
each solution and then average the result to search for an
asymmetry. We have checked with Monte Carlo events that this process
does not bias our result for the case of no {\em CP\/} violation. For
the case of {\em CP\/} violation, we use a special Monte Carlo
calibration procedure described below.

We use a data set that corresponds to $13.3\,{\rm fb^{-1}}$ of total
integrated luminosity and contains 12.2 million $\tau^+\tau^-$
pairs produced from $e^+e^-$ collisions at a center-of-mass energy
near or at the $\Upsilon(4S)$. We select events for which each $\tau$
decays via the mode $\tau\rightarrow\pi\pi^0\nu$, selected for its
large branching fraction ($\simeq 25\%)$) and high selection efficiency
(10\%). We verify with Monte Carlo events that our selection criteria
do not introduce a bias in the $\xi$ distribution.

The major source of background are $\tau$-pair events where one of the
$\tau$'s decays to a $\rho\nu$ final state, the other $\tau$ decays to
a $\pi\pi^0\nu$ final state, and one of the $\pi^0$'s is not
reconstructed. Using Monte Carlo events, we estimate this contamination
level at 5.2\% of the selected data. The overall background
contamination from $\tau$ decays is estimated at 10\%.  We estimate
that events of the form $e^-e^+\rightarrow q\bar{q}$ ($q= u,d.c,s$
quarks), $e^-e^+\rightarrow \Upsilon(4{\rm S})\rightarrow B\bar{B}$,
and $e^-e^+\rightarrow e^-e^+\gamma\gamma$ contribute a background
contamination of less than 0.1\%.

\subsection{Calibration}

\begin{figure}
\center
\psfig{figure=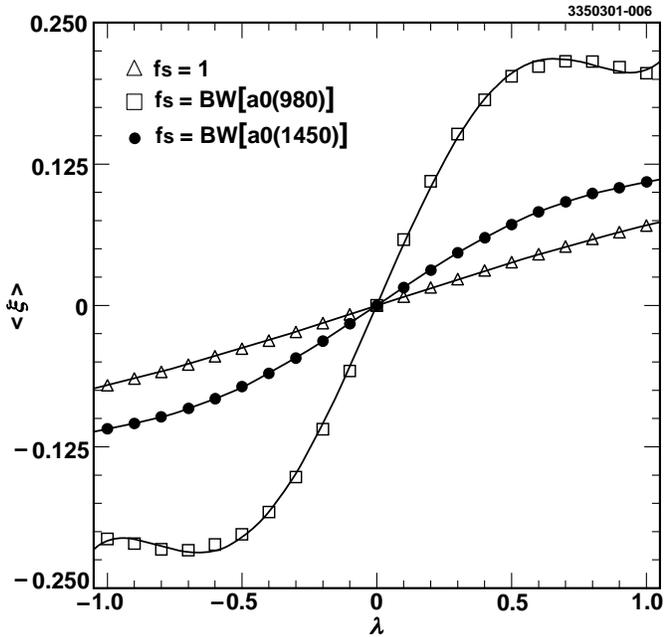,width=3.50in}
\caption{$\langle\xi\rangle$ v. $\lambda$ for each of the three choices
for the scalar form factor: a) ${\rm f_s=1}$, b) ${\rm f_s= BW[a_0(980)]}$, 
and c) ${\rm f_s= BW[a_0(980)]}$.  }
\label{fig:lambda}
\end{figure}

From Eq.~(\ref{eq:lambda}) we see that to lowest order
$\langle\xi\rangle=c_1\lambda$.  To determine or limit $\lambda$, we
need to measure $\langle\xi\rangle$ and determine the calibration
constant $c_1$. This is done in a two step process. First, we
determine the range over which $\langle\xi\rangle$ is a linear
function of $\lambda$, using Monte Carlo events and assuming perfect
detector response. Figure~\ref{fig:lambda} shows $\langle\xi\rangle$
v. $\lambda$ for the three choices of the scalar form factor. After
this linear range is determined, we choose  5 values of $\lambda$ within
that range, and with full detector simulation compute
$\langle\xi\rangle$ for each of the 5 $\lambda$ values. This produces
a set of 5 ($\lambda, \langle\xi\rangle$) points to which a straight
line is fit.  The resulting slope is then $c_1$. This entire process
is done for each of the 3 scalar form factors.

\section{Results}
\subsection {Observed $\xi$ Distributions}

\begin{figure}[hhh]
\center
\psfig{figure=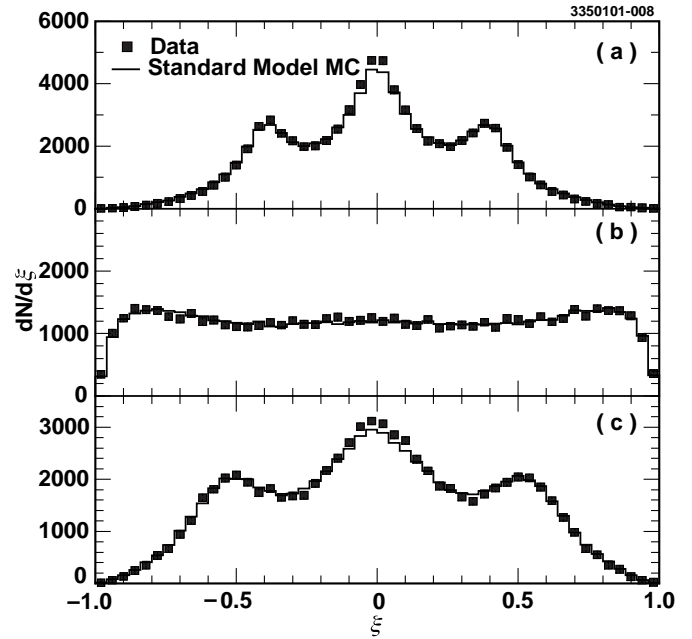,width=3.50in}
\caption{The distribution for the {\em CP}-sensitive observable $\xi$
for each of the three choices for the scalar form factor: a) ${\rm f_s = 1}$,
b) ${\rm f_s = BW[a_0(980)]}$, and c) ${\rm f_s = BW[a_0(1450)]}$.
The squares are the data and the line is SM Monte Carlo.}
\label{fig:xires}
\end{figure}

Using the calibration constant, the $\xi$ distribution for data and
Monte Carlo is shown in figure~\ref{fig:xires} for each of the scalar
form factors. The data are squares and the line is SM Monte Carlo. The
corresponding measurements for $\langle\xi\rangle$ and the 90\% CL
limits on $\lambda$, the imaginary part of the non-SM coupling, are
shown in table~\ref{tab:xilim}.

\begin{table}[htbp]
\begin{center}
\caption{Limits on $Im(\Lambda)$ for various
 form factors $f_s$.}\label{tab:xilim}
\begin{tabular}{|c|c|c|}\hline 
${\rm f_s}$ & $\langle\xi\rangle\times 10^{-3}$
& $Im(\Lambda) \ @ \ $ 90\% CL \\
\hline


${\rm f_s = 1}$ & $-0.8\pm1.4$ & $-0.046< \lambda < .022 $\\
${\rm f_s = BW[a_0(980)]}$     & $-0.1\pm0.4$ &$ -0.008< \lambda < .006 $\\
${\rm f_s = BW[a_0(1450)]}$   & $0.1\pm1.2$ &$ -0.019< \lambda < .021 $\\

\hline
\end{tabular}
\end{center}
\end{table}

\subsection{Systematics}

\begin{figure}[hhh]
\center
\psfig{figure=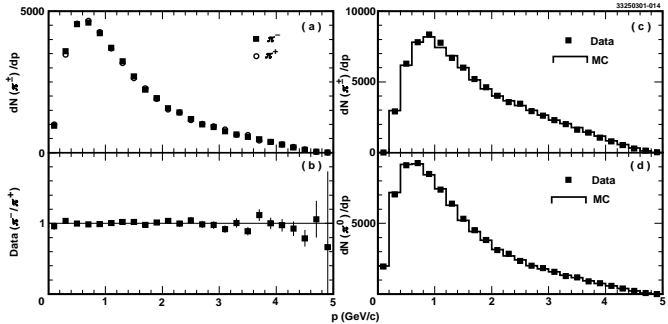,width=3.50in}
\caption{The $\pi^{\pm}$ momentum distributions for data are shown
in a) while the ratio is shown in b). The distributions for data and
Monte Carlo are shown for charged pions c) and for neutral pions
d).}
\label{fig:pip}
\end{figure}

Care is taken to verify that the detector does not create an
artificial $\xi$ asymmetry. Systematic effects due to a possible
difference between track reconstruction efficiencies for $\pi^-$ and
$\pi^+$ as a function of pion momentum are studied. The momentum
distribution for charged pions in the reactions
$\tau^{\pm}\rightarrow\pi^{\pm}\pi^0\nu_{\tau}$ is plotted in
figure~\ref{fig:pip}a. The corresponding ratio of these distributions
is shown in figure~\ref{fig:pip}b and is seen to be consistent with
1. Varying the slope of this ratio by $\pm1\sigma$ leaves $c_1$
unchanged but does change $\lambda$ by $\pm0.003$. We take this as a
measure of the systematic error due to differences in the charged
pion track reconstruction efficiencies.

Imperfect Monte Carlo simulation of the momentum distributions for the
tau decay products can also bias our results. The momentum distributions
for charged and neutral pions for both data and Monte Carlo are
shown in figure~\ref{fig:pip}c and figure~\ref{fig:pip}d. The agreement
is quite good. Varying the slope of these distributions by $\pm1\sigma$
permits an estimate of the systematic error on $\lambda$ to be
made. The effect is negligible ($\sim10^{-5}$).

A possible asymmetry in the $\xi$ distribution due to background
events has been estimated with Monte Carlo and found to be negligible.

\begin{figure}
\center
\psfig{figure=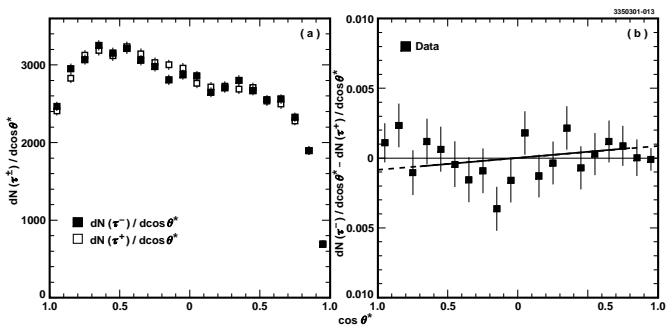,width=3.50in}
\caption{The pseudo-helicity distribution for charged
taus is shown in a) while their difference
is shown in b).}
\label{fig:scalar}
\end{figure}

\section{Search for Scalar Mediated $\tau$-Decays}

We also search in a model independent way and without using the
optimal observable $\xi$ for the presence of scalar mediated tau
decays with the same final state as before. In the SM, the helicity
angle $\theta_{\pi\pi^0}$, defined as the angle between the direction of
the charges pion in the $\pi\pi^0$ rest frame and the direction of the
$\pi\pi^0$ system in the $\tau$ rest frame, has a distribution
appropriate to the exchange of a vector particle:

\be
{dN \over {d\,\cos\theta_{\pi\pi^0}}} \propto a + b\cos^2\theta_{\pi\pi^0}. 
\ee

Including a non-SM scalar exchange leads to a $\cos\theta_{\pi\pi^0}$
term in the distribution, corresponding to S-P wave interference
that is proportional to the non-SM complex coupling $\Lambda$:

\be
{dN \over {d\,\cos\theta_{\pi\pi^0}}} \propto a
+ ( c_1Re(\Lambda) + c_2 Im(\Lambda))\cos\theta_{\pi\pi^0}
+ b\cos^2\theta_{\pi\pi^0}.
\label{eq:sp}
\ee

\noindent Observation of a term proportional to
$\cos\theta_{\pi\pi^0}$ would imply {\em CP\/} violation. However, the
precise direction of the taus is unknown, due to the
presence of a final state neutrinos, so the distribution in
Eq.~\ref{eq:sp} cannot be formed. Instead, we use the
`pseudo-helicity' angle $\theta^{\ast}$, defined as the angle between
the charged pion in the $\pi\pi^0$ rest frame and the direction of the
$\pi\pi^0$ system in the lab frame.  This doesn't change the form of
Eq.~\ref{eq:sp}, just the coefficients.

The term in Eq.~\ref{eq:sp} containing $Im(\Lambda)$ changes sign for
taus of opposite signs.  Using the angle $\theta^{\ast}$, the difference
between pseudo-helicity distributions for oppositely charged
taus is given by

\be
{dN(\tau^-) \over {d\,\cos\theta^{\ast}}}
- {dN(\tau^+) \over {d\,\cos\theta^{\ast}}}
\sim 2c_2Im(\Lambda)\cos\theta^{\ast}.
\label{eq:pseudo}
\ee

\noindent The presence of a $\cos\theta^{\ast}$ term in such a
distribution implies {\em CP\/} violation.

The data sample used for this search is the same as before.
The pseudo-helicity distribution for charged taus is shown in
figure~\ref{fig:scalar}a. The structure in the distributions is due to
the variation in detection efficiency as a function of $\pi^{\pm}$
momentum and $\pi^0$ energy.  To measure or constrain $Im(\Lambda)$ we
need to determine the proportionality constant $c_2$. This is done with
a procedure similar to the one used for the 3HDM case.

The difference in pseudo-helicity distributions is shown
in figure~\ref{fig:scalar}b. The slope extracted from
the data distribution is $c_2Im(\Lambda)=(4.2\pm3.6)\times 10^{-4}$,
consistent with zero. This leads to a constraint on $Im(\Lambda)$:

\be
-0.033< Im(\Lambda) < 0.089 \ @ \ 90\%\ {\rm CL}
\ee

\noindent Note that this result is consistent with, but less
restrictive than our conservative limit shown in the first line of
table~\ref{tab:xilim}. This is due to the use of a non-optimal {\em
CP}-sensitive variable.






\section*{Acknowledgments}
I thank the organizers for an excellent conference and,
in particular, A. Sanda for permitting this talk on such short notice.

\end{document}